\documentstyle[preprint,prl,aps,epsfig]{revtex}
\begin{document}
\def\sn2{$\sin^22\theta$}
\def\dm2{$\Delta m^2$}
\def\ch2{$\chi^2$}
\def\n{$\nu~$}
\def\nue{$\nu_e~$}
\def\nus{$\nu_s~$}
\def\numu{$\nu_{\mu}~$}
\def\nutau{$\nu_{\tau}~$}
\def\ar{$\rightarrow~$}
\def\lrar{$\leftrightarrow~$}
\def\ltap{\ \raisebox{-.4ex}{\rlap{$\sim$}} \raisebox{.4ex}{$<$}\ }
\def\gtap{\ \raisebox{-.4ex}{\rlap{$\sim$}} \raisebox{.4ex}{$>$}\ }
\draft
\begin{titlepage}
\preprint{\vbox{\baselineskip 10pt{
\hbox{IC/98/30}
\hbox{ }
\hbox{}}}}
\vskip -0.4cm
\title{ \bf Parametric Resonance in  Oscillations of 
Atmospheric Neutrinos?}
\author{Q.Y. Liu$^{1)},~$ S.P. Mikheyev$^{2,4)},~$  A.Yu. Smirnov$^{3,4)}$
\footnote{ e-mail addresses:  qyliu@sissa.it,  Stanislav@lngs.infn.it,
Smirnov@ictp.trieste.it.}}
\address{1) Scuola Internazionale Superiore di Studi Avanzati, I-34013
Trieste, Italy}
 
\address{2) Laboratory Nazionale del Gran Sasso dell'INFN,
I-67010 Assergi (L'Aquilla), Italy} 

\address{3) Abdus Salam International Centre for Theoretical Physics,
I-34100
Trieste, Italy}
\address{4) Institute for Nuclear Research, Russian Academy of Sciences, 
107370 Moscow, Russia}
\maketitle
\begin{abstract}
\begin{minipage}{5in}
\baselineskip 16pt
We consider a solution of the atmospheric neutrino problem based on 
oscillations of muon neutrinos to sterile neutrinos: 
$\nu_{\mu}$ $\leftrightarrow$ $\nu_s$. 
The zenith angle ($\Theta$) dependences of the neutrino and 
upward-going muon fluxes in presence of these
 oscillations are studied. The dependences 
have characteristic form with two dips:   
at $\cos \Theta = -0.6 \div -0.2$
and $\cos \Theta = -1.0 \div -0.8$. The latter dip is due to parametric
resonance in oscillations of neutrinos which cross the core of the
earth. A comparison of predictions with data from the MACRO, Baksan and
Super-Kamiokande experiments is given.

\end{minipage}
\end{abstract}
\end{titlepage}
\newpage


\def\C{C}
\def\S{S}

\setcounter{footnote}{0}


\vglue 1.5cm
\leftline{\bf 1. Introduction}
\vskip 0.2cm
Recently, the Super-Kamiokande \cite{SK} 
and  Soudan \cite{Soudan} experiments have further confirmed 
an existence of the atmospheric neutrino problem
\cite{Kamiokande,IMB}.  Moreover, 
observations of the zenith angle  
as well as $L/E$ (distance/energy) 
dependences of the muon neutrino 
deficit strongly indicate an oscillation 
solution of the problem.


Oscillations should give an observable effect in the zenith angle 
($\Theta$-)dependence of the flux of the upward-going muons 
produced by high energy neutrinos. 
However, the experimental situation is not yet clear.
A peculiar $\Theta$-dependence has been observed by MACRO experiment 
\cite{MACRO}. 
The ratio of the
observed and expected fluxes as the function of 
$\Theta$  has two dips. The wide
dip is  in the range
$\cos \Theta = -0.6~\div~-0.2$ with minimum at $\cos \Theta =
-0.4~\div~-0.5$, and  the narrow dip is at $\cos \Theta =
-1.0~\div~-0.8$ (vertical bins). In the narrow dip the observed flux is
almost two times smaller than the expected one. 
Between these two dips  ($\cos \Theta \sim -0.8$) 
the flux is close to  the value without oscillations. 
It is important that for the vertical bins 
there is the best understanding
of the acceptance and efficiencies. The probability of 
the no oscillation hypothesis is smaller than 1$\%$, and inclusion
of $\nu_{\mu}$ $\rightarrow$ $\nu_{\tau}$ oscillations does not raise
the probability in a significant way \cite{MACRO}.
Baksan experiment \cite{BAKSAN} has similar threshold for muon detection. 
The data also show a wide dip at $\cos \Theta =
-0.6~\div~-0.2$ similar to that in MACRO data. However, no
significant deficit has been observed in the vertical bins. The
Super-Kamiokande \cite{SK} with higher threshold than in MACRO and
Baksan shows less profound  
$\Theta$ - effect, although some suppression of the flux is observed both at 
$\cos \Theta = -0.6~\div~-0.2$ and in the vertical bins.

In \cite{LS} it was observed that oscillations of muon neutrinos to
sterile neutrinos, $\nu_{\mu}$ $\leftrightarrow$ $\nu_s$, lead to the
zenith angle dependence being qualitatively similar to that observed by
MACRO experiment. In particular, it has been shown that the narrow 
dip in vertical bins can be due to the parametric enhancement 
of oscillations
for neutrinos whose trajectories cross both the mantle and the core of
the Earth. The oscillation effect decreases with increase of the neutrino
energy. This can explain absence of strong deficit in 
experiments with high thresholds like the Super-Kamiokande.

The $\nu_{\mu}$ $\leftrightarrow$ $\nu_s$ oscillations give as good
description of the low energy (sub-GeV and multi-GeV) data as $\nu_{\mu}$
$\leftrightarrow$ $\nu_{\tau}$ do (see e.g. \cite{FootVolYas}). The
appropriate region of oscillation parameters 
\begin{equation} 
\Delta m^2~=~(2 \div 10) \cdot 10^{-3}~\rm{eV}^2,~~~~~~\sin^2
2\theta ~>~0.8~ 
\end{equation} 
is shifted slightly to
larger values of $\Delta m^2$ in comparison with the $\nu_{\mu}$
$\rightarrow$ $\nu_{\tau}$ case
due to  matter effect which becomes important for
$\nu_{\mu}$ $\leftrightarrow$ $\nu_s$ channel at small $\Delta m^2$.

Crucial difference between $\nu_{\mu} \leftrightarrow \nu_s$ 
and $\nu_{\mu} \leftrightarrow \nu_{\tau}$
channels appears for high energy
atmospheric neutrinos which are detected by upward-going muons. The
matter effect is absent in $\nu_{\mu}$ $\leftrightarrow$
$\nu_{\tau}$ channel but it is  important for 
$\nu_{\mu} \leftrightarrow \nu_s$ \cite{Akhmedov}.

In this paper we perform a detailed study of the zenith angle
dependences of the atmospheric neutrino and muon fluxes
in presence of the $\nu_{\mu}$ $\leftrightarrow$ $\nu_s$ oscillations. 
We give a description of the parametric resonance effect 
in sect. 2. 
In sect. 3 we present results of
calculations of the zenith angle dependences for the through-going
and stopping muons. In sect. 4 we compare the prediction with
experimental results.     


\vskip 0.3cm
\leftline{\bf 2. Parametric resonance in atmospheric neutrinos}
\vskip 0.2cm

The $\nu_{\mu}$ $\leftrightarrow$ $\nu_s$ oscillations in matter are
described by the $2 \times 2$ evolution matrix (the effective Hamiltonian)
with off-diagonal elements $H_{\mu s}=H_{s \mu}=\Delta m^2 / 4E \cdot 
\sin 2\theta$ and the difference of the diagonal elements: 
\begin{equation}
H_{\mu \mu} - H_{ss}~=~\displaystyle { {\Delta m^2} \over {2E}}  \cos
2\theta~- ~V~.
\label{ham}
\end{equation}   
Here $\theta$ is the vacuum mixing angle, $\Delta m^2 \equiv m^2_2 -
m^2_1$ is the mass squared difference 
($m_2$ is the mass of the neutrino  component which has 
larger admixture in $\nu_{\mu}$ in the case of non-maximal mixing), $E$
is the energy of neutrino and
$V$ is the matter potential
\begin{equation} 
V~=  \frac{G_F}{\sqrt{2}} \displaystyle {   N_n~   }~,  
\label{poten}
\end{equation} 
where $G_F$ is the Fermi constant, $N_n$ is the neutron number density
\footnote{Notice that in general case of $\nu_{\mu}$ mixed with
$\nu_{\tau}$ and $\nu_e$, the off-diagonal (mixing) elements get
contributions from interaction with matter,
and also the potential depends on electron/proton density.}.
(For antineutrinos $V$ has an opposite sign).

According to (\ref{ham}, \ref{poten}) the length 
of $\nu_{\mu} \leftrightarrow \nu_s$ oscillations 
in matter and the effective mixing angle $\theta_m$
are determined by 
\begin{equation} 
l_m~=~\frac{2\pi}{V}\left[ (\xi \cos 2\theta + 1)^2 + \xi^2
\sin^2 2\theta  \right]^{-1/2},
\label{length}
\end{equation}
\begin{equation} 
\sin^2 2\theta_m~=~\sin^2 2\theta ~\xi^2 \left[ 
(\xi  \cos 2\theta + 1)^2 + \xi^2 \sin^2 2\theta  \right]^{-1},
\label{mixing}
\end{equation}
where 
\begin{equation} 
\xi ~=~\frac{ \Delta m^2}{2EV}~.
\label{xi}
\end{equation} 
In the case of maximal mixing the
equations (\ref{length}) and
(\ref{mixing}) are simplified:
$$
l_m = \frac{2\pi}{V}\left[ 1 + \xi^2  \right]^{-1/2}, ~~~ 
\sin^2 2\theta_m = \frac{\xi^2}{1 + \xi^2}~~.  
$$

For $\Delta m^2> 4 \cdot 10^{-3}$ eV$^2$ both for 
the sub-GeV and multi-GeV ($E \sim 3 \div 5$ GeV) 
events  we get $\xi^2 > 10$ and the 
matter effect is small ($ < 10 \%$ in probability).  
Thus, in significant region of parameters 
(masses and mixing) at low energies 
the $\nu_{\mu} \leftrightarrow \nu_s$ oscillations reproduce  
results of the  
$\nu_{\mu} \leftrightarrow \nu_{\tau}$ oscillations, 
as far as the charged current interactions are concerned. 
(The number of the tau-lepton events is small.)

The matter effect  becomes important
for multi-GeV events if  $\Delta m^2< 3 \cdot 10^{-3}$ eV$^2$
\cite{LS,FootVolYas}.   
In particular,  the zenith angle dependence of ratios is modified. 
Since  matter suppresses the $\nu_{\mu} \leftrightarrow \nu_s$ 
oscillations, one expects weaker oscillation effect for the 
upward-going neutrinos, and therefore, 
flatter   zenith angle dependence than
in the $\nu_{\mu}$ $\leftrightarrow$ $\nu_{\tau}$ case where matter
effect is absent.

The matter effect is important for distributions of the 
through-going and stopping muons \cite{Akhmedov} 
produced by high energy neutrinos  
even for large $\Delta m^2$. 

Let us first describe the zenith angle dependence of the survival
probability  $P(\nu_{\mu}$ $\rightarrow$ $\nu_{\mu})$ for fixed
neutrino energy. To a good approximation the Earth can be considered
as consisting of the core and the mantle with 
constant densities and sharp change of
density at the border between the mantle and the core. Therefore 
propagation of neutrinos through the Earth has a character of
oscillations in layers with constant densities. 
This gives not only correct qualitative picture but also rather precise
(as we will see) quantitative results.

According to (\ref{length}) 
 with increasing  energy (decreasing  $\Delta m^2$) $\xi
\rightarrow 0$ 
and the oscillation length increases approaching the 
asymptotic value  determined by 
the potential only: $l_m \approx 2 \pi/V $. 
At the same time, the effective mixing angle decreases 
(\ref{mixing}), so that 
the oscillation effects become weaker. 
Let us consider 
the range of energies which  corresponds to
$\xi = 0.1 \div 1$. 
Here $l_m$ weakly
depends on $\xi$, but $\sin^22\theta_m$ is not yet
strongly suppressed.  
In this case the phase of oscillations, $\Phi$, acquired by neutrino 
in a given layer $\Delta L$ equals:

\begin{equation} 
\Phi = 2 \pi \int^{\Delta L} \frac{dL}{l_m} \approx \int^{\Delta L} 
 d L ~ V ~.  
\label{phase}
\end{equation} 
The phase $\Phi$  depends on density and  size of the layer,
and it does not depend on neutrino energy.
Using this feature we can immediately get the zenith angle dependence of 
$P$($\nu_{\mu}$ $\rightarrow$ $\nu_{\mu}$). Notice that for $\cos \Theta =
-0.8$ neutrino trajectories touch the core of the Earth. Therefore
for $\cos \Theta > -0.8$ neutrinos cross the mantle only,  
 whereas for $\cos \Theta \lesssim -0.8$
they cross both the mantle and the core. 

For $\cos \Theta \geq -0.8$ (mantle effect only), the survival
probability can be written as 
\begin{equation}	
P \approx 1 - \sin^2 2\theta_m \sin^2 \frac{\Phi_m (\Theta)}{2},  
\label{start1}
\end{equation}	
where the phase acquired by neutrinos, $\Phi_m (\Theta)$,  equals: 
\begin{equation} 
\displaystyle { \Phi_m (\Theta) \approx  2R_E  |\cos \Theta| V
~\approx \sqrt{2} G_F N_n
 R_E |\cos \Theta|}~. 
\label{phasem}
\end{equation} 
Here $R_E$ is the radius of the Earth. From (\ref{phasem}) we find  
$\Phi(\cos \Theta = -0.4) = \pi$, so that
maximal oscillation
effect, $P_{max}=1-\sin^2 2\theta_m$, is at $\cos \Theta = -0.4$. At $\cos
\Theta = -0.8$  the phase is $\Phi_m = 2\pi$, and the oscillation effect is
zero.

For $\cos \Theta < -0.8$, neutrinos cross three layers: mantle, core
and again mantle. The survival probability is 
$P=|S_{\mu \mu}|^2$, where the 
evolution matrix $S$ equals 
\begin{equation} 
S~=~U(\theta_m) D(\Phi_m) U^+(\theta_m-\theta_c)
 D(\Phi_c) U(\theta_m-\theta_c)  D (\Phi_m) U^+(\theta_m)~.
\label{Smatrix}
\end{equation}
Here $U(\theta_m)$ is the $2 \times 2$ mixing matrix, $\theta_m$ and
$\theta_c$ are
the mixing angles in matter of the mantle and core correspondingly. The matrix 
\begin{equation}
D(\Phi)~=~diag(1, \displaystyle { e^{i \Phi}}) 
\label{Dmatrix}
\end{equation} 
describes the evolution of the eigenstates in certain layer; 
$\Phi_c$ and $\Phi_m$ are the phases of oscillations acquired 
in the core and in each layer of the mantle. 
It turns out that $\Phi_m \approx \pi$. Moreover, for
$\cos \Theta \approx -0.88$ also $\Phi_c$ equals $\pi$. Thus for 
$\cos \Theta \approx -0.88$ we get the equality:
\begin{equation}
\Phi_{m1} \approx \Phi_{c} \approx \Phi_{m2} \approx \pi~.  
\label{equality}
\end{equation} 
Under  this condition an enhancement of oscillations occurs. Indeed, now
$D=diag(1,-1)$ for all layers, and inserting this $D$ in (\ref{Smatrix}),
we get the  probability
\begin{equation} 
P  = |S_{\mu \mu}|^2 = 1 -  \sin^2 (4 \theta_m - 2\theta_c).  
\label{probab}
\end{equation}
Clearly for $\theta_c < \theta_m$, the probability 
$P$ is smaller than  $1 - \sin^2 2\theta_m$ and  $1 - \sin^2
2\theta_c$ corresponding  to maximal
oscillation effect in one density layer. 
Moreover, the smaller $\theta_c$ (the stronger suppression of mixing
in the core) the bigger the transition effect.

The condition
(\ref{equality}) means that the size of the layer, R, coincides with
half of the oscillation length: $R=l_m /2$. This is the
condition of the parametric resonance \cite{param,param2}
which can be written in general as,
$2\pi \int^{r_f}_0 dr /{l_m} ~=~2\pi k$,  $k=1,2,3...$, 
where $r_f$ is the size (period) of perturbation,  and in our case $r_f =
R_c + R_m$.    
Graphical representation \cite{param2} of the enhancement is shown in fig. 1:
Mixing angle changes suddenly when the   
phase of system reaches $\pi$. This leads to increase of the
oscillation angle.
Thus one expects the (parametric resonance) peak at $\cos \Theta
= -0.88$. The width of the peak is $\Delta \cos \Theta \approx 0.12$.
For exact vertical direction $\cos \Theta = -1~$, we get 
$\Phi_c \approx 2.5 \pi$   
and the enhancement is destroyed. Thus 
unsuppressed flux is expected in  this direction.

Let us stress that for sufficiently large energies 
the equality (\ref{equality}) does not 
depend on neutrino masses. The equality  is determined basically 
by the density distribution in the Earth and by the  
potential which, in turn, is fixed by the channel of oscillations 
and the Standard Model interactions.  
The equality is fulfilled for oscillations into 
sterile neutrinos only. In  the $\nu_{\mu} \leftrightarrow \nu_{e}$
channel the  
potential is two times larger and equality (\ref{equality}) is 
broken.

With the increase of the neutrino energy the form of the zenith angle
dependence (position of minima and maxima) practically does 
not change, however
the mixing angle, and therefore the depth of oscillations, 
diminish. Oscillation effect disappears  for $\xi \ll 0.1$.  In
contrast, for low energies, $\xi \geq 1$, it reduces to the vacuum
oscillations effect. Thus the profound zenith angle dependence 
with two dips exists in  rather narrow range:
$
\xi ~\approx~0.1~\div ~0.5
$
(fig. 2). 
This dependence
coincides qualitatively with the zenith angle dependence of the muon flux
observed by the MACRO experiment.  
The $\Theta$-distributions in fig. 2 have been calculated for 
real density profile of the Earth \cite{earth}. 
They close to the results
 obtained from (\ref{start1}) (\ref{Smatrix}) for simplified model of the Earth (layers with constant density)  
(see fig.2 in \cite{LS}).  


\vskip 0.3cm 
\noindent
{\bf 3. Zenith angle distribution of the upward-going muons} 
\vskip 0.2cm 

The flux of muons with energy above the threshold energy $E^{th}_{\mu}$ as a
function of the zenith angle equals
\begin{equation} 
F_{\mu}(E_{\mu}^{th}, \Theta)~=~\int_{E_{\mu}^{th}} dE \cdot 
\sum_{ i=\nu , \bar{\nu}}  F_i(E ,\Theta) \cdot P_i(E, \Theta) \cdot 
Y_i(E,E_{\mu}^{th}),
\label{FluxD}
\end{equation}
where $F_{\nu}(E, \Theta)$ is the flux of original neutrinos,
$P(E, \Theta)$ is the survival probability, and 
\begin{equation} 
Y_i(E,E_{\mu}^{th})~=~\int_{E_{\mu}^{th}} R(E_{\mu}^{\prime},E_{\mu}^{th}) \cdot  
\frac{d\sigma_i}{d E_{\mu}^{\prime}} (E, E_{\mu}^{\prime})~
\end{equation}
is the probability that the neutrino with energy $E$ produces muon, which arrives at a
detector with the energy above $E_{\mu}^{th}$. 
Here $R(E_{\mu}^{\prime},E_{\mu}^{th})$ is the muon range (the
distance in $g/cm^2$ at which muon energy decreases from $E'_{\mu}$ to
$E^{th}_{\mu}$ due to energy loss) and 
$d \sigma / d E_{\mu}^{\prime}$ is the differential cross section
of $\nu_{\mu}N$ $\rightarrow$ $\mu X$ reaction.

The integrand in (\ref{FluxD}) at $P=1$:
\begin{equation} 
I~=~F(E ,\Theta) \cdot Y(E,E_{\mu}^{th})
\label{spectrum}
\end{equation}
describes the energy distribution of neutrinos that give rise to muon
flux above $E^{th}_{\mu}$. It has the form of a wide peak 
with full width on the half of height characterized by 
$1.5 \div 2$ orders of magnitude in $E$.
This is substantially bigger than the
width of region with significant two dips effect 
($\xi = 0.1 \div 0.5$). A position of maximum of the peak,
$E_{\nu}^{max}$, depends on $E_{\mu}^{th}$ and $\Theta$, and for
$E_{\mu}^{th}  \sim 1$ GeV  
one has
$E_{\nu}^{max} \approx (20 \div 30)$ GeV.

Due to integration over 
neutrino energies the $\Theta$ - distribution of muons differs from  the 
distribution of neutrinos shown in fig. 2.  
There are two modifications:

(i) Smoothing of $\Theta$ dependence occurs due to contributions from low
energy neutrinos ($\xi \ge 1$) which undergo basically averaged
vacuum oscillations effect. 
This leads to decrease of the peak between two dips in
$\Theta$-dependence. 

(ii) Weakening of the  overall suppression of the flux
occurs  due to contribution from high energy neutrinos 
with $\xi < 0.1$. For these
neutrinos the oscillation effect is small due to smallness of
mixing angle.

The dependence of the ratio of fluxes with and without oscillations 
$F_{\mu} /F^0_{\mu}$ on $\Theta$ for different values
of $\Delta m^2$ is shown in fig. 3a. With increase of $\Delta m^2$
the region of strong oscillation effect ($P(E)$) shifts to higher
energies and thus larger part of the neutrino spectrum will undergo
suppression. This leads to overall strengthening of the oscillation
effect, however the contribution from low energies also increases and
the two dips zenith angle dependence becomes less profound. In contrast, for small
$\Delta m^2$ the $\Theta$-dependence is more profound, but overall
suppression is weaker. Similar changes occur when the threshold,
$E_{\mu}^{th}$, varies (fig. 3b). With increase of  $E_{\mu}^{th}$
(for fixed $\Delta m^2$) the peak of the integrand (\ref{spectrum}) 
shifts to  higher energies; 
the $\Theta$-dependence with two dips becomes more profound but overall
suppression effect weakens. 
 This can explain 
weaker effect in experiments with higher $E_{\mu}^{th}$ (like
the Super-Kamiokande). (Note that there is no
${E^{th}_{\mu}}/ {\Delta m^2}$ scaling: the change of $\Delta m^2$
leads to stronger effect than similar change of $1/E^{th}_{\mu}$.)

For non-maximal mixing a dependence of the oscillation length on the
neutrino energy becomes stronger: it appears in the lowest order in
$\xi$.  Indeed, from (\ref{length}) we have  
$$
l_m~\approx~\frac{2\pi}{V} \frac{1} {1 \mp \xi \cdot \cos 2\theta}~,
$$
where for  $\Delta m^2 > 0$ 
minus and plus signs are  for  neutrinos and antineutrinos 
correspondingly. 
For non-maximal mixing the oscillation length of neutrinos is larger
and  the condition of the parametric
resonance is fulfilled for larger $|\cos\Theta|$. Thus, the parametric
peak shifts to vertical direction. For antineutrinos the length is
smaller
and the effect is opposite. Since the ${\nu}$-flux is
larger than $\bar{\nu}$-flux the parametric peak for total flux 
shifts to larger   $|\cos\Theta|$ (fig. 3c).

Uncertainties in the 
shape of the neutrino energy spectrum (and cross sections) 
can influence the zenith angle dependence. 
To study this effect we parameterize possible modifications 
of the spectrum, and consequently the integrand $I$ as
\begin{equation}
I(E) = I_0 (E) A E^{\alpha + \beta \ln (E/1{\rm GeV})},  
\label{alphabeta}
\end{equation} 
where $I_0(E)$ is the integrand which corresponds to the
spectrum from  \cite{Agrawal}, $\alpha$ and $\beta$ 
are parameters and $A$
is a normalization factor. 
The results of the calculations for different values of parameters 
 $\alpha$ and $\beta$ are shown in fig. 3d. As follows from fig. 3, 
shift of the maximum to higher energies leads to suppression of the 
vacuum oscillation effect and therefore to more profound
zenith angle dependence (dashed line), however the overall suppression
becomes weaker. In contrast, the shift of the maximum to lower energies 
enhances the vacuum oscillation effect (dashed dotted line). 
Both more profound zenith angle dependence  and strong overall suppression
can be obtained for narrower $I(E)$ distribution (dotted line).

For stopping muon sample, the peak
$I_{stop} (E)$ is narrow
and its maximum is at smaller energies: $E \sim 10$ GeV. For
$\Delta m^2 \geq 5 \cdot 10^{-3}$ eV$^2$ the matter effect is small ($\leq
10\%$), so that the zenith angle distribution is determined basically by
vacuum oscillations (fig. 4). Difference between $\nu_{\mu}$ $\rightarrow$
$\nu_s$ and $\nu_{\mu}$ $\rightarrow$ $\nu_{\tau}$ channels is small. 
For $\Delta m^2 \leq 4 \cdot 10^{-3}$ eV$^2$ the matter effect becomes
important. The zenith angle dependence acquires the form with two dips.

Similarly, the sample of muons produced in the detector itself 
corresponds to low energies and narrow peak of the intergand,  thus
leading essentially to vacuum oscillations effect. So, we expect 
strong suppression of  
the number of these events which weakly depends on zenith angle. 


\vskip 0.3cm
\noindent
{\bf 4. Predictions versus experimental results}
\vskip 0.2cm

Let us compare the predicted distributions with the experimental  
results  from the Baksan, MACRO (fig.5a) and Super-Kamiokande experiments
(fig. 5 b).
Histograms show the $\Theta$-distribution averaged over  the bins 
for ``probe'' point
$\Delta m^2 = 8 \cdot 10^{-3}$ eV$^2$, $\sin^2 2\theta=1$, and
neutrino spectrum from \cite{Agrawal}.
The predicted suppression in the vertical bin is still
weaker than in MACRO but
stronger than in Baksan data. The predicted curve would fit better
average of MACRO and Baksan results.
Integrations over neutrino and muon energies result in
strong smoothing of the distribution.
The difference of suppressions in the dips and in
the region between the dips is about  (15 $\div$ 20)
$\%$. Without parametric effect one would expect the ratio
$F_{\mu} / F_{\mu}^0 \rightarrow 1$ in the vertical bins.
Notice that still error bars are
large and the experimental situation for vertical bins is
controversial (fig. 5a). In the same time
there is a good agreement of predictions
 with the Super-Kamiokande zenith angle 
dependence (fig. 5b). Clearly, more
data are needed to make any conclusion.

As follows from fig. 3, varying the neutrino parameters
($\Delta m^2$, $\sin^2 2\theta$)
and modifying neutrino spectrum one can change
the predicted $\Theta$-dependence. 
In particular, it is possible to make the dips move profound, to
change relative depth of the dips, to shift positions of minima in
$\cos\Theta$. 
Fitting the data one can also take into
account  uncertainties in normalization
of original neutrino flux  which can reach
$20\%$.

Clearly $\Theta$- dependences for $\nu_{\mu}$ $\rightarrow$ $\nu_s$ and
$\nu_{\mu}$ $\rightarrow$ $\nu_{\tau}$ channels are different. The
 $\nu_{\mu}$ $\rightarrow$ $\nu_{\tau}$ oscillations lead to smooth
enhancement of  suppression with $|\cos \Theta|$, whereas the
 $\nu_{\mu}$ $\rightarrow$ $\nu_s$ oscillations result in the
structure with two dips.
Furthermore, for vertical bins a suppression due to  $\nu_{\mu}$
$\rightarrow$  $\nu_{\tau}$
is stronger than due to  $\nu_{\mu}$ $\rightarrow$
$\nu_s$ and  the difference can be as big as $20\%$.
For $\nu_{\mu}$ $\rightarrow$ $\nu_{\tau}$ channel
suppression weakly depends on the energy threshold,
so that for Baksan/MACRO and Super-Kamiokande experiments
the dependences are similar. In contrast, for
$\nu_{\mu}$ $\rightarrow$ $\nu_s$ channel the suppression effect
decreases with increase of the threshold. 
The effect is large for stopping
muons (or muons produced in the detector), it is weaker 
for through-going muons in
Baksan/MACRO and the weakest effect is for through-going muons 
in the Super-Kamiokande.

The two channels can be distinguished also by studying neutral current
effects \cite{VS}.

In conclusion,
in the range of $\nu_{\mu}$ $\rightarrow$ $\nu_s$ oscillations parameters, which correspond to the best fit
of the low energy data, the zenith angle dependence of the
upward-going muons has a peculiar
form with two dips at $\cos \Theta = (-0.6 \div
-0.2)$ and  $\cos \Theta = (-1.0 \div -0.8)$. The dip in vertical
direction is due to parametric resonance in oscillations of neutrinos 
which cross the core of the Earth. This zenith angle dependence is
crucial signature of a solution of the atmospheric neutrino problem, based on $\nu_{\mu}$ $\leftrightarrow$
$\nu_s$ oscillations.

\vskip 0.3cm
\leftline{\bf Acknowledgments.} 
The authors are grateful to E. Kh. Akhmedov and  P. Lipari for
useful discussions. 
The work of Q.Y.L. is
supported in part by the EEC grant ERBFMRXCT960090.

\newpage 

\noindent
{\bf Figure Captions}\\

\noindent
Fig. 1. Graphical representation of the parametric enhancement of
neutrino oscillations. Vector $\overrightarrow{\nu}$ describes the
neutrino state in such a way that its projection on the axis $Y$ gives $2P-1$,
where $P$ is the
probability to find $\nu_{\mu}$-neutrino. (The axis $X$ is
Re($\psi^*_{\mu} \psi_s$), where $\psi_{\mu}$ and $\psi_s$ are the wave
functions of the muon and sterile neutrinos.) The positions of
$\nu$-vector 2 - 4 correspond to $\nu$-states
at the borders of mantle and the core.

\noindent
Fig. 2. The zenith angle dependence of the survival probability $P($
$\nu_{\mu}$ $\rightarrow$ $\nu_{\mu}$ $)$ for different 
neutrino energies: solid line - 100 GeV, dashed line - 50 GeV, dash-dotted
line - 30 GeV ($\Delta m^2 = 8 \cdot 10^{-3}$ eV$^2$).

\noindent
Fig. 3. The zenith angle dependence of the ratio $F_{\mu} / F_{\mu}^0$
for the upward-going muons. 
a). the $\Theta$ dependences for different values of $\Delta m^2$ 
($\sin^2 2\theta=1$  and  $E_{\mu}^{th} = 1$ GeV);
b). the same for different energy thresholds of muon detection,
$E_{\mu}^{th}$ ($\sin^2 2\theta=1$ and $\Delta m^2 = 8 \cdot 10^{-3}$ eV$^2$); 
c). the same for different values of mixing angle  
($\Delta m^2 = 8 \cdot 10^{-3}$
eV$^2$ and  $E_{\mu}^{th} = 1$ GeV); 
d). the same for different original spectra of neutrinos described by
parameters $\alpha,~\beta$ in (\ref{alphabeta}): 
$\alpha$ = 0.75, $\beta$ = -0.075 (dashed line) 
$\alpha$ = 0.25, $\beta$ = -0.075 (dash-dotted line) 
$\alpha$ = 0.75, $\beta$ = -0.10 (dotted line), 
solid line corresponds to  unmodified  neutrino flux.

\noindent
Fig. 4. The zenith angle dependence of the ratio $F_{\mu} / F_{\mu}^0$
for stopping muons for different values of $\Delta m^2$ 
($\sin^2 2\theta=1$ and  $E_{\mu}^{th} = 1$ GeV). 
   
\noindent
Fig. 5. Comparison of the predicted zenith angle dependence
(histograms) with
experimental data from the MACRO and Baksan (a) and Super-Kamiokande
($E_{\mu}^{th}= 7$ GeV) (b) experiments.
Solid lines for $\nu_{\mu}$ $\rightarrow$ $\nu_s$ oscillations with
$\Delta m^2=8 \cdot 10^{-3}$ eV$^2$ and $\sin^2 2\theta=1$; dashed
line for $\nu_{\mu}$ $\rightarrow$ $\nu_{\tau}$ oscillations with
$\Delta m^2=5 \cdot 10^{-3}$ eV$^2$ and $\sin^2 2\theta=1$.

\begin{figure}[H]
\vglue 1.8cm
\mbox{\epsfig{figure=fig1.eps,width=14cm,height=11cm,angle=0}}
\vglue2.5cm
\caption[]{ }
\end{figure}

\begin{figure}[H]
\vglue 1.8cm
\mbox{\epsfig{figure=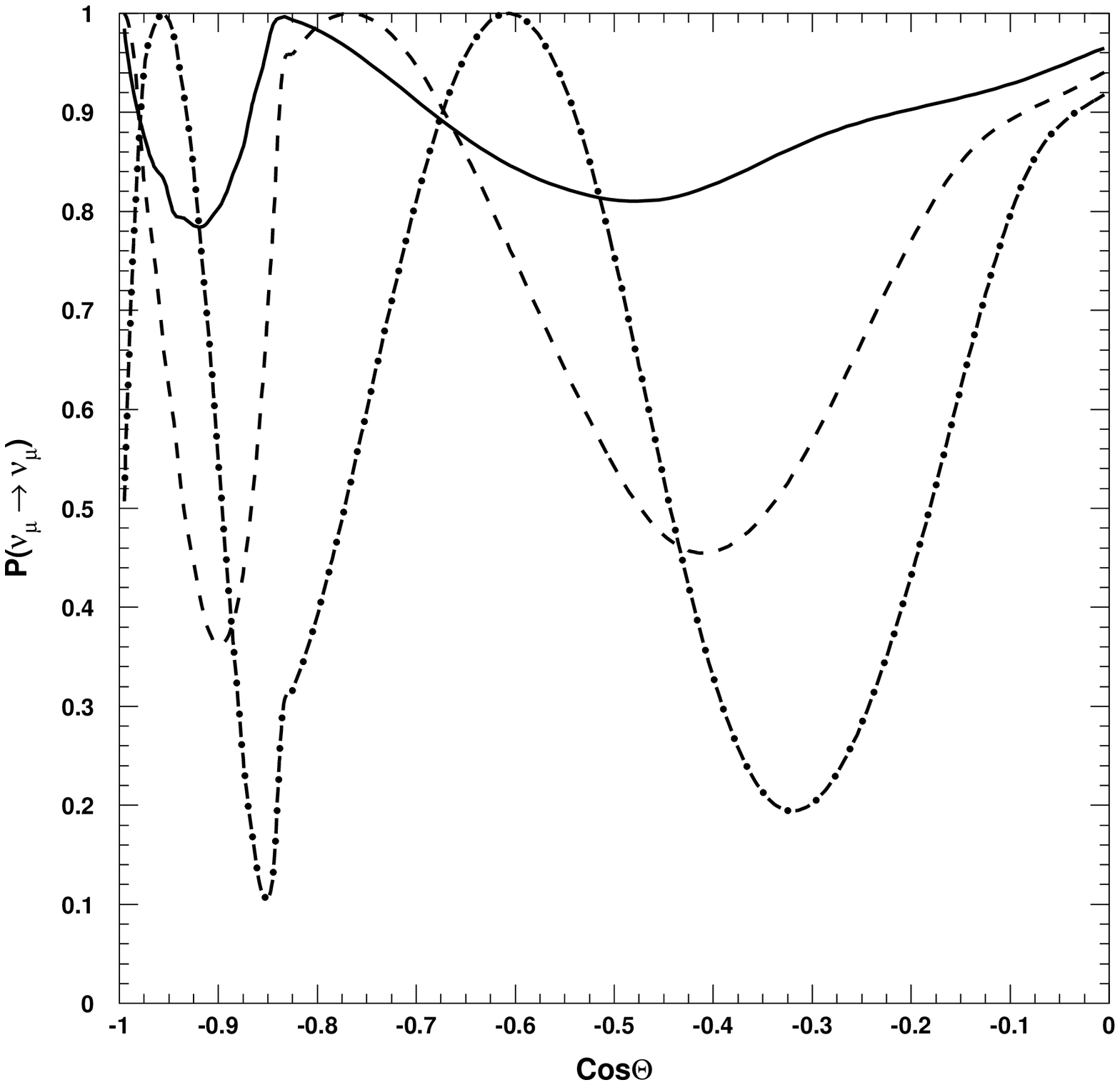,width=14cm,height=11cm,angle=0}}
\vglue2.5cm
\caption[]{ }
\end{figure}

\begin{figure}[H]
\mbox{\epsfig{figure=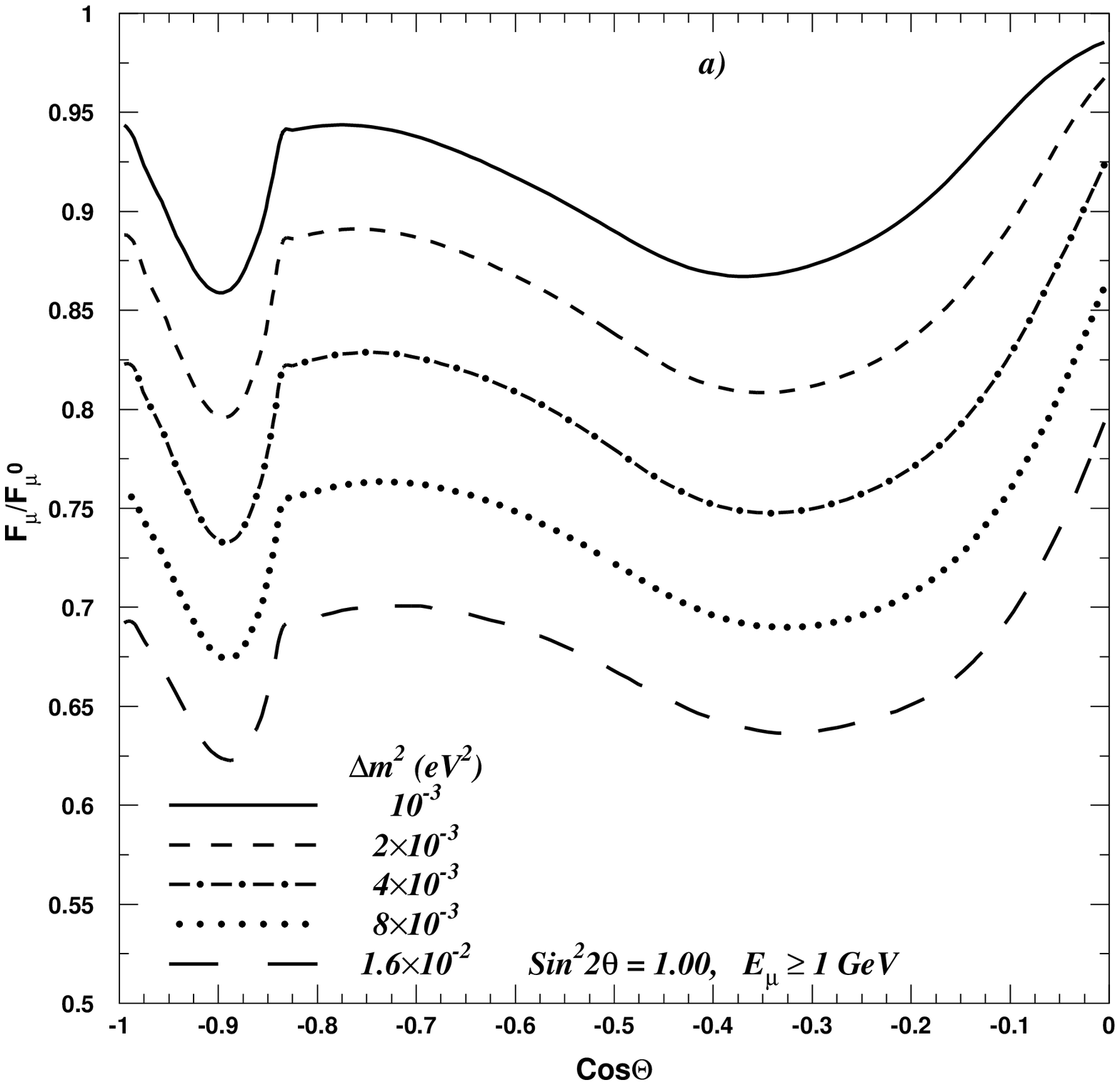,width=14cm,height=11cm,angle=0}}
\vglue 0.1cm
\indent \mbox{\epsfig{figure=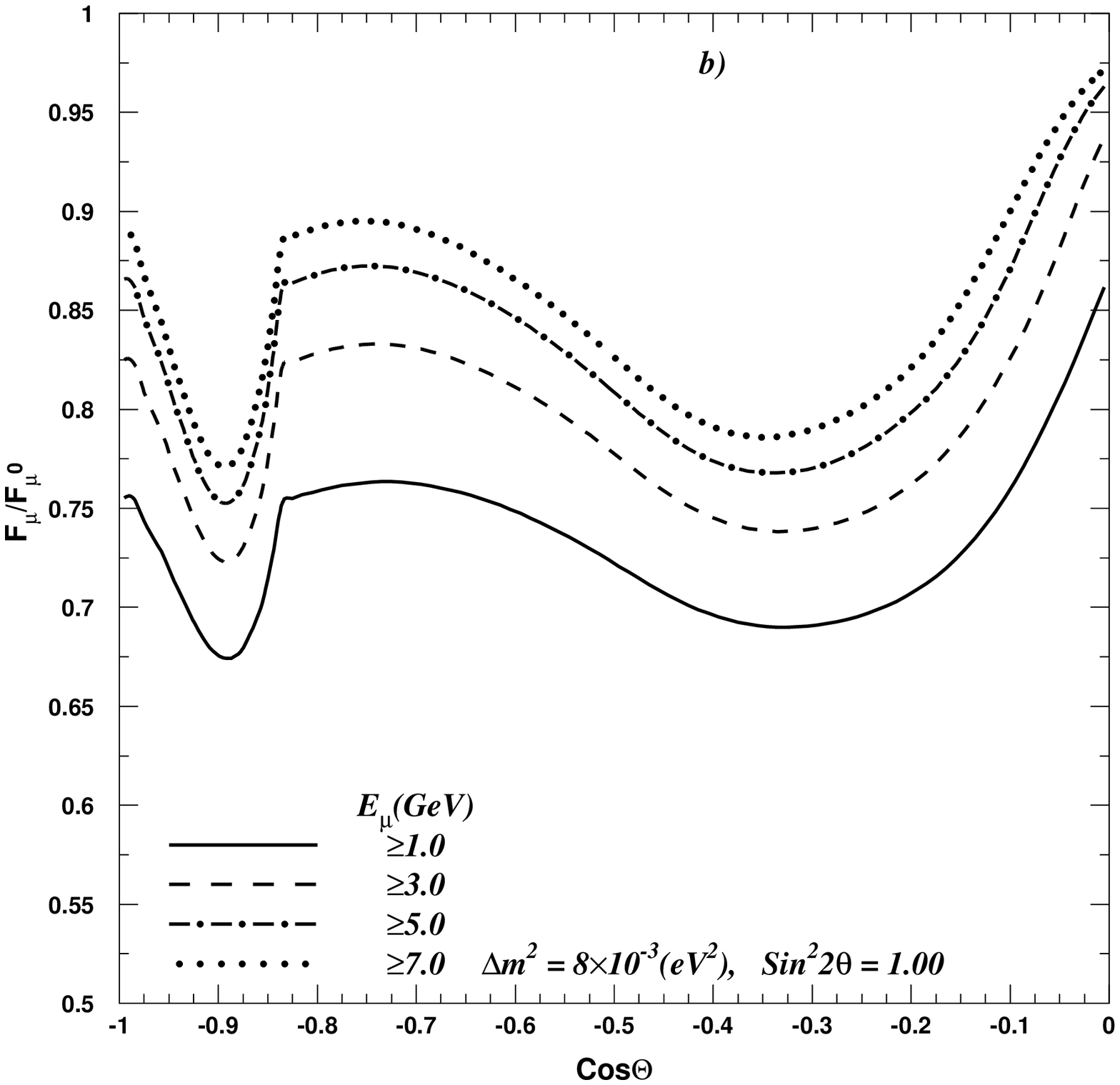,width=14cm,height=11cm,angle=0}}

\newpage
\mbox{\epsfig{figure=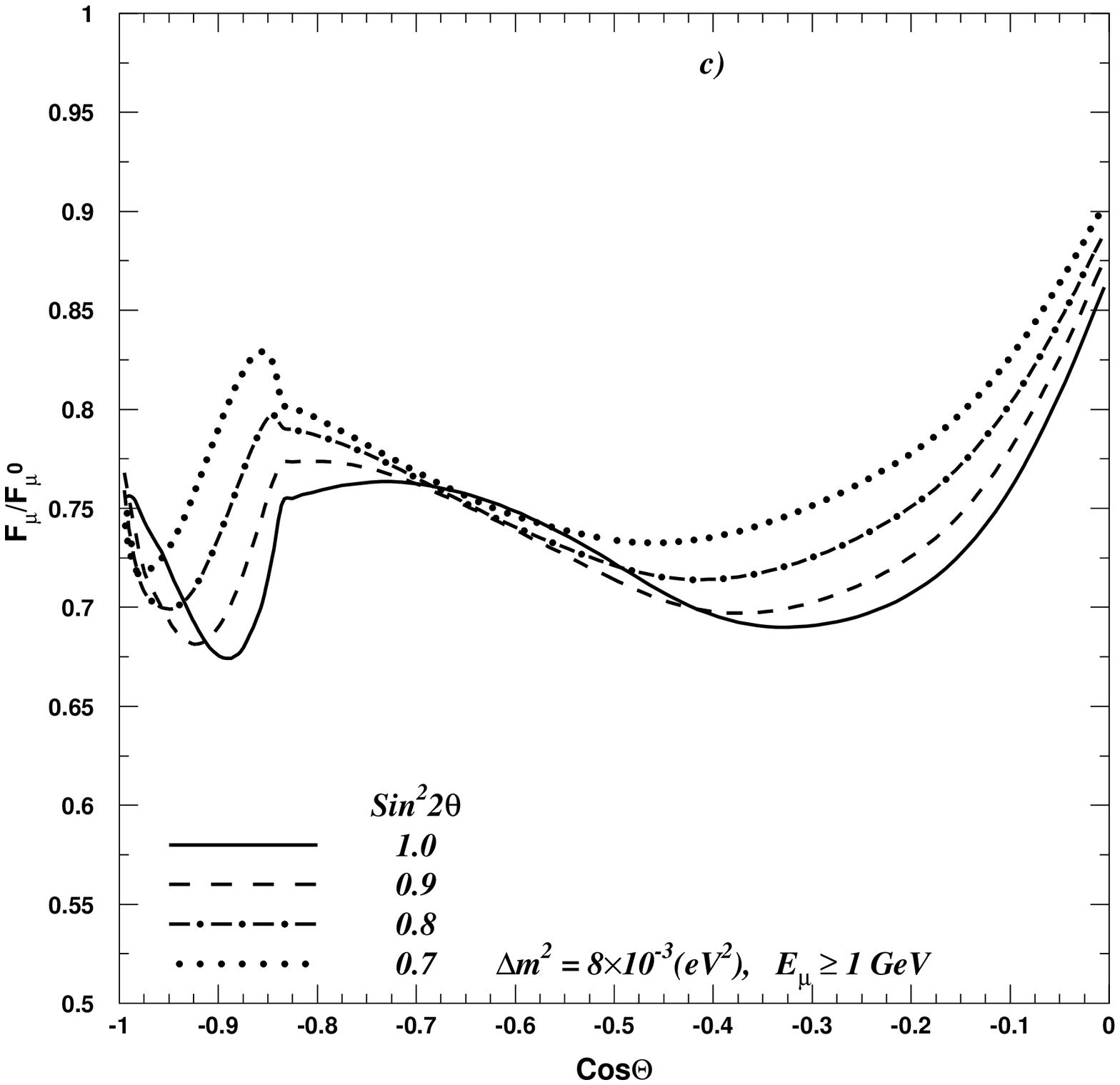,width=14cm,height=11cm,angle=0}}
\indent \mbox{\epsfig{figure=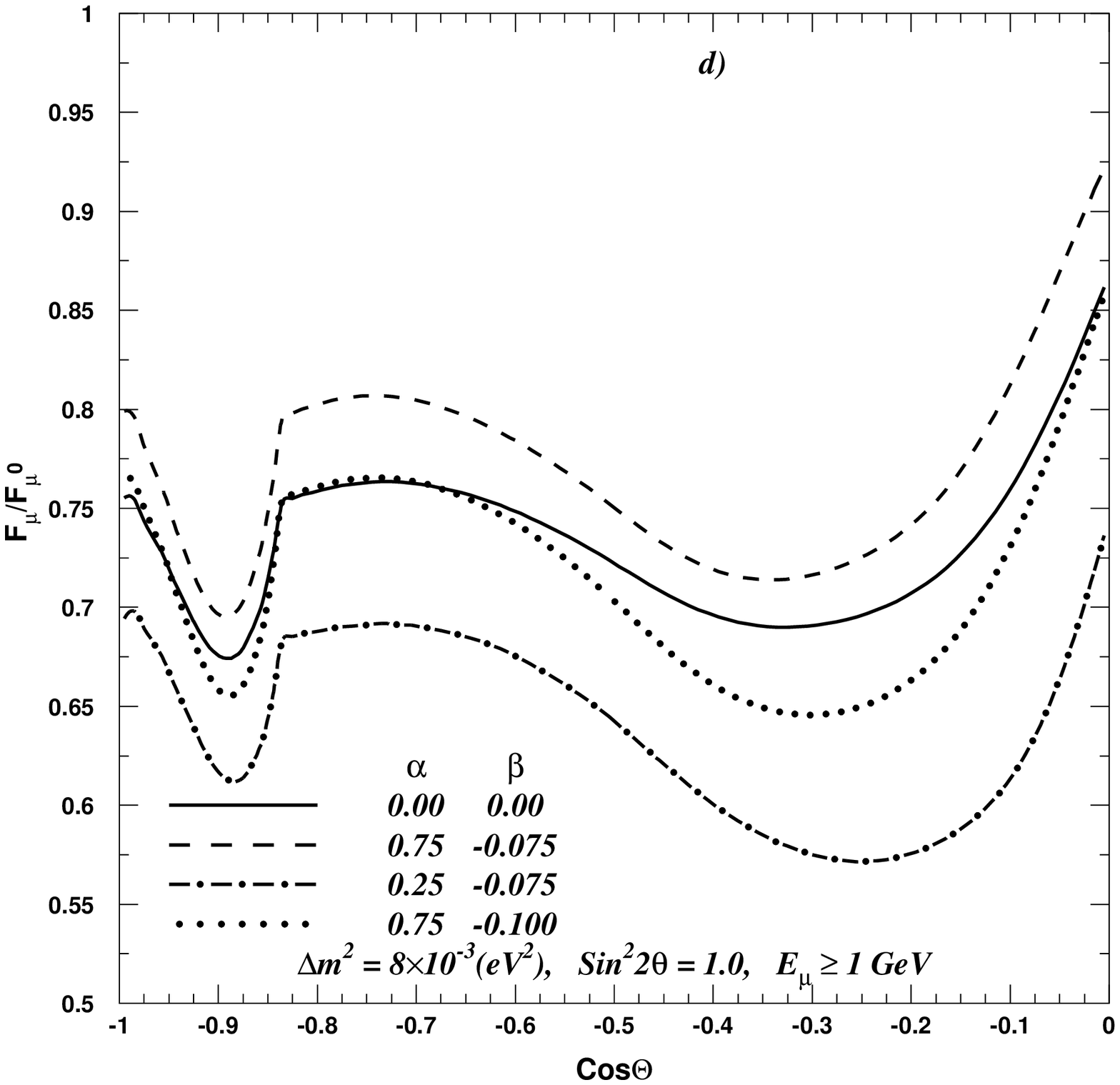,width=14cm,height=11cm,angle=0}}
\caption[]{ }
\end{figure}

\begin{figure}[H]
\vglue 1.8cm
\mbox{\epsfig{figure=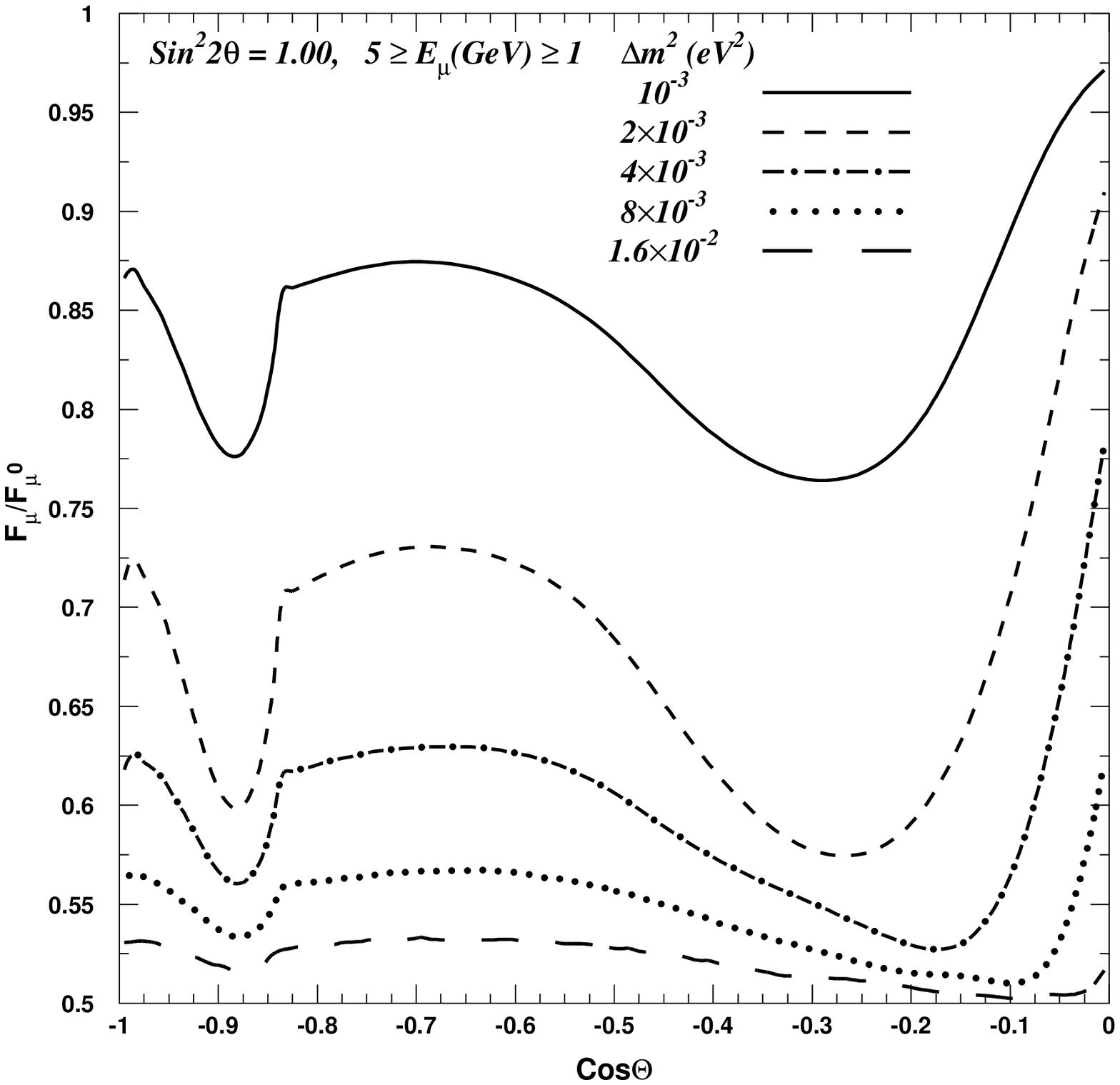,width=14cm,height=11cm,angle=0}}
\vglue2.5cm
\caption[]{ }
\end{figure}

\begin{figure}[H]
\mbox{\epsfig{figure=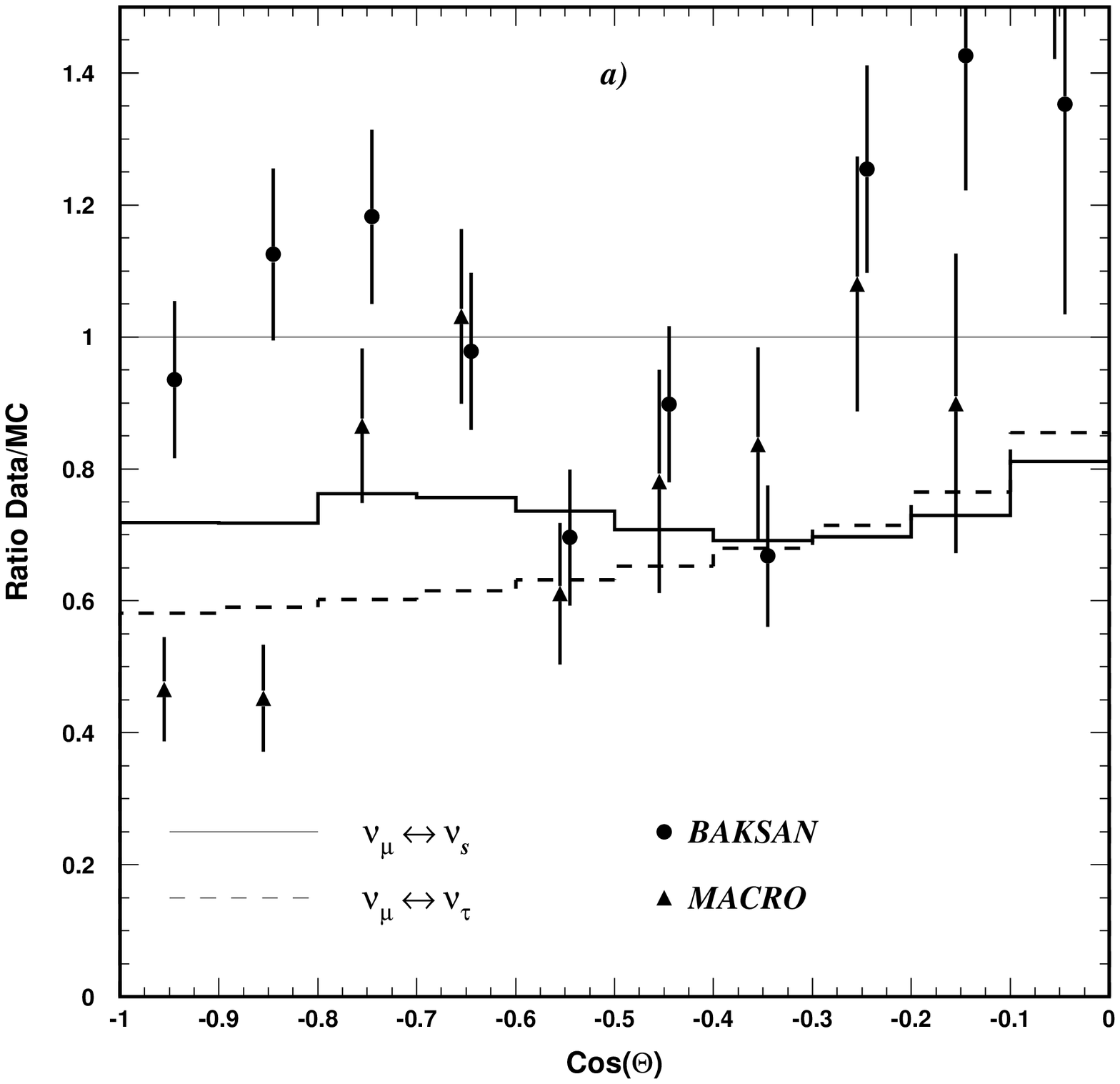,width=14cm,height=11cm,angle=0}}
\indent \mbox{\epsfig{figure=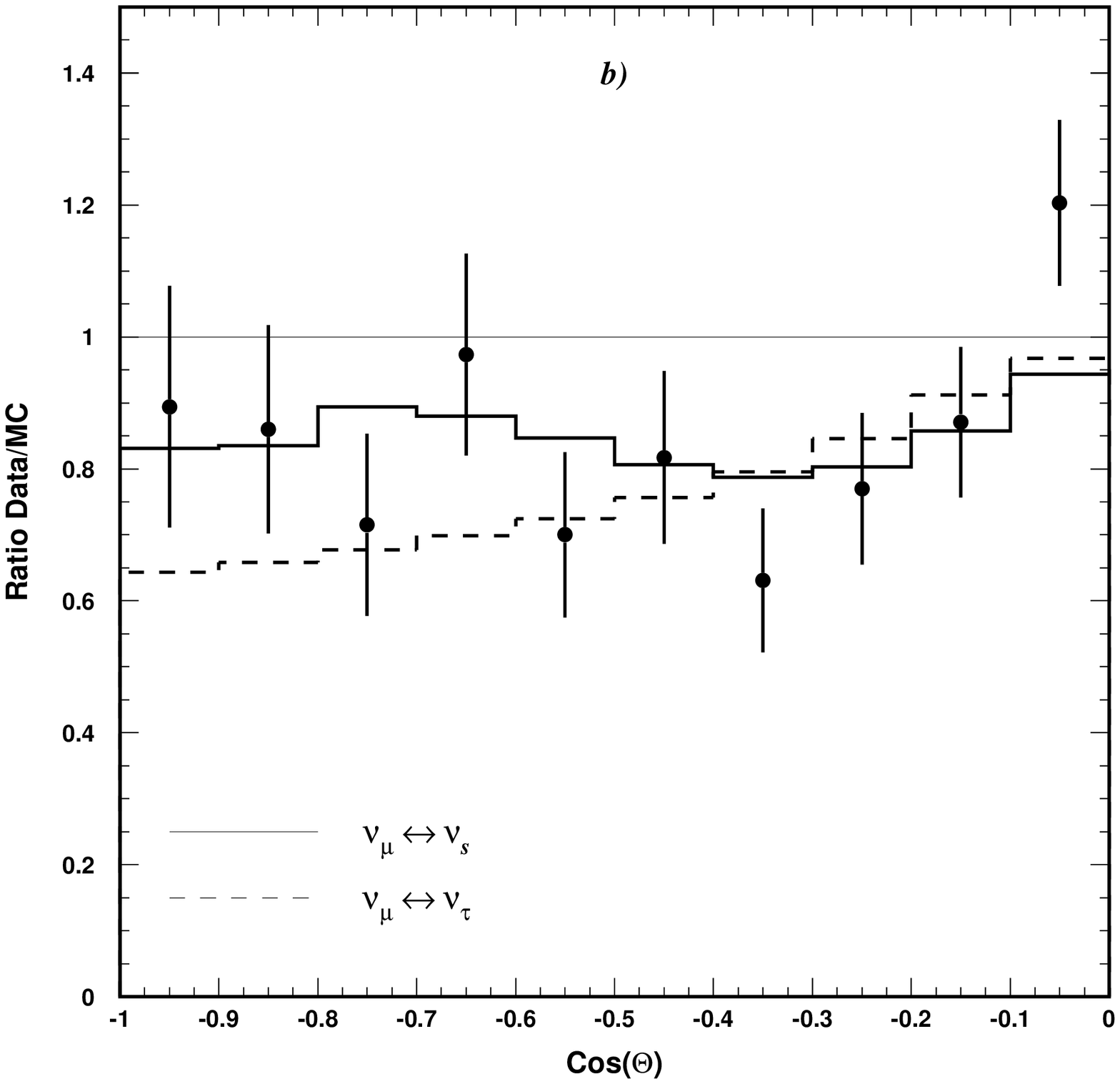,width=14cm,height=11cm,angle=0}}
\caption[]{ }
\end{figure}




\begin{thebibliography}{99}



\bibitem{SK} Y.Totsuka (Super-Kamiokande Collaboration), in $LP'97$,
28th Int. Symposium on Lepton Photon Interactions, Hamburg,
Germany, 1997, to appear in the Proceedings;\\
E. Kearns, talk given at conference on $Solar~Neutrinos:~News~About~SNUs$,
December 2-6, 1997, Santa Babara.



\bibitem{Soudan} S.M. Kasahara et al., 
Phys. Rev. {\bf D 55} (1997) 5282; T. Kafka, talk given at Fifth
Int. Workshop TAUP-97, September 7 - 11, 1997, Gran Sasso, Italy.


\bibitem{Kamiokande} Y. Fukuda et al.,  Phys. Lett. {\bf B 335} (1994) 237.


\bibitem{IMB} R. Becker-Szendy 
et al., Phys. Rev. {\bf D 46}, (1992) 3720; Nucl. Phys. B (Proc. Suppl.)
{\bf 38} (1995) 331. 


\bibitem{MACRO} F. Ronga, Proceedings of the 17th International
Conference on Neutrino Physics and Astrophysics ($Neutrino ~96$), 
(1996) 529, edited by
K. Enqvist, K. Huitu and J. Maalampi; 
T.Montaruli, talk given at TAUP 97, Sep. 1997 (Gran Sasso, Italy).


\bibitem{BAKSAN} M.M. Boliev, et al., Proc. of the 8th ``Rencontres de
Blois'' 
($Neutrinos,~dark~matter~and~the~universe$), edited by T. Stolarcyk,
J. Tran Thanh Van, F. Vannucci (1996) 296. 


\bibitem{LS} Q. Y. Liu and A. Yu. Smirnov, hep-ph/9712493.


\bibitem{FootVolYas} R. Foot, R.R. Volkas, O. Yasuda, TMUP-HEL-9801, 
 (hep-ph/9801431); TMUP-HEL-9803, (hep-ph/9802287).
 




\bibitem{Akhmedov} E. Akhmedov, P. Lipari, M. Lusignoli, 
Phys. Lett. {\bf B 300} (1993) 128. 





\bibitem{param} V.K. Ermilova, V.A. Tsarev and V.A. Chechin, Kr. Soob,
Fiz. Lebedev Institute 5 (1986) 26; 
E. Akhmedov, Yad. Fiz. {\bf 47} (1988) 475 (Sov. J. Nucl. Phys. 
{\bf 47} (1988) 301). 

\bibitem{param2} P. I. Krastev and A. Yu. Smirnov, Phys. Lett. {\bf
B 226} (1989) 341. 


\bibitem{earth} see {\it e.g.} F. D. Stacey, 
{\it Physics of the Earth} (John Wiley and Sons, New York, 
$2^{nd}$ edition, 1977. 



\bibitem{Agrawal} V.Agrawal et al., Phys. Rev. D{\bf 53} (1996) 1314.





\bibitem{VS} F. Vissani and A.Yu. Smirnov, hep-ph/9710565.


 
\end{thebibliography}
\end{document}